\documentclass{article}
\usepackage{graphicx}
\usepackage{amsmath}
\usepackage{amsfonts}
\usepackage{amssymb}

\begin{document}

\title{On Interpretation of Special Relativity:\\a complement to\\Covariant Harmonic Oscillator Picture}
\author{Paul Korbel\\Institute of Physics, \\University of Technology, ul. Podchorazych 1,\\30-084 Cracow, Poland}
\maketitle
\begin{abstract}
In 1971 Feynman, Kislinger and Ravndal \cite{Fosc} proposed Lorentz-invariant
differential equation capable to describe relativistic particle with mass and
internal space-time structure. By making use of new variables that
differentiate between space-time particle position and its space-time
separations, one finds this wave equation to become separable and providing
the two kinds of solutions endowed with different physical meanings. The first
kind constitutes the running waves that represent Klein-Gordon-like particle.
The second kind, widely discussed by Kim and Noz \cite{Kbook}, constitutes
standing waves which are normalizable space-time wave functions. To fully
appreciate how valuable theses solutions are it seems necessarily, however, to
verify a general outlook on relativity issue that (still) is in force. It was
explained \cite{PK267} that Lorentz symmetry should be perceived rather as the
symmetry of preferred frame quantum description (based on the freedom of
choice of comparison scale) than classical Galilean idea realized in a
generalized form. Currently we point to some basic consequences that relate to
solutions of Feynman equation framed in the new approach. In particular (i)
Lorentz symmetry group appears to describe energy-dependent geometry of
extended quantum objects instead of relativity of space and time measure, (ii)
a new picture of particle-wave duality involving running and standing waves
emerges, (iii) space-time localized quantum states are shown to provide a new
way of description of particle kinematics, and (iv) proposed by Witten
\cite{Witten} generalized form of Heisenberg uncertainty relation is derived
and shown be the integral part of overall non-orthodox approach.
\end{abstract}

\newpage

\section{Introduction}

Since early seventies Kim and Noz \cite{Kim73} proposed unconventional outlook
on the issue of relationship between the quantum mechanics and special
relativity. Although they did not admit this openly, in fact, they have
suggested that special relativity can be understood properly only within the
framework of quantum mechanics. In many of his papers Kim pointed out that
known aspect of length contraction wins its clarify meaning exclusively
against a background of wave description. Such a view allows us to avoid many
confusing and misleading interpretations that plague relativity from its very
beginning and which cannot be removed if one relays exclusively on the
classical approach. As noticed by Kim: ``If not possible, it is very difficult
to formulate Lorentz boosts for rigid bodies. On the other hand, it seems to
be feasible to boost waves'' \cite{Kim04}.

The author of this paper strongly support the thesis that special relativity
is integral part of quantum mechanics and that separation between these two
realms is apparent and artificial. Simple analysis given in \cite{PK267}
showed that the source of Lorenz symmetry is not relativity of inertial frames
but the freedom of choice of comparison scale imposed on two quantities being
given in different physical units (like energy and momentum, and/or distance
and time). In particular it was shown that idea of comparison scale combined
with two fundamental postulates of quantum mechanics, of \ Planck and de
Broglie, provides the basis of covariant relativistic description, thereby
predicts basic dynamical features of relativistic particles. It was explained
also that Lorentz symmetry needs to be seen as the symmetry of preferred frame
(i.e. observer rest frame) quantum description. In this paper we continue the
progress along that line by indicating that Lorentz symmetry group is a
natural tool to describe internal space-time structure of extended quantum
objects. Discussed in the paper the main consequences resulting from such
non-orthodox point of view are:

1. The time and length measure relativity aspect is taken over by issue of
energy-dependent space-time deformation of extended quantum objects.

2. Space-time localized quantum states are shown to provide a new way of
description of particle kinematics.

3. A new picture of particle-wave duality involving running and standing waves
becomes visible.

The solutions of covariant harmonic oscillator equation provided by Kim and
Noz \cite{Kbook} state plausible illustration for the presented ideas.

The structure of the paper is the following: In \textbf{Section 2} we start
with a pure quantum-mechanical analysis intended to recollect the way which
two light-cone momenta put together may constitute a four-momentum of
relativistic particle with mass. Then we introduce a concept of light-cone
skeleton, which will enable us to link particle dynamical features with its
space-time extensions and thus with its kinematical properties. The
differential equation of Feynman, Kislinger and Ravndal \cite{Fosc}, i.e. the
covariant harmonic oscillator equation, as well as, the basic properties of
solutions of this equation obtained by Kim and Noz are discussed in
\textbf{Section 3}. We show/recollect also the way which scalar particle gains
its mass due to vibrations of its internal space-time structure and analyze
the particle ground state geometry. The light-cone skeleton structure,
introduced in Section 2, turns out to be a useful quantum-mechanical tool to
characterize excitation levels of \textit{extended oscillatory} particle. The
two kinds of such excitations, called respectively \textit{kinetic }and
\textit{potential} are distinguished and described in \textbf{Section 4}. Both
of them are shown to be space-time localized but their physical meanings
differ. The kinetic excitations are identified with four-momentum transfer
carried out by running plane-waves embraced, however, within a space-time
localized area which extensions are determined just by the light-cone skeleton
structure. On the other hand, the same light-cone structure is shown to spread
out the potential excitations formed by standing waves. A new picture of
particle-wave duality expressed in terms of kinetic and potential excitations
is the subject of \textbf{Section 5.} The issue of \textit{oscillatory motion}
of extended quantum object is discussed next. The energy and momentum
fluctuations that relate to this peculiar kind of motion are shown to be in
Heisenberg-Witten relation which is a generalized form of Heisenberg
uncertainty relation proposed by Witten \cite{Witten}. Finally, in
\textbf{Section 6}, we recollect some experimental results that in spectacular
manner disclose non-locality of quantum mechanics.

\section{Quantum-mechanical description of relativistic particle with mass}

The Klein-Gordon equation for scalar particle or the Dirac equation for
spin-half particle are the basic tools of relativistic quantum field theory
that make descriptions of massive particles possible. Nevertheless, the
introduction of particle mass in the relativistic case, similarly like in the
non-relativistic case of Schr\"{o}dinger quantum mechanics, is equivalent to
introduction of mass parameter. Since each of the mentioned wave equations
cannot predict itself possible particle internal space-time structure, it is
widely believed that point-particle picture is the most accurate one.

Conventional field theory methods, however, indicates also another, quite
different way of mass introduction. A very instructive example provides us the
solutions of covariant harmonic oscillator equation given by Kim and Noz
\cite{Kbook}. Their analysis is complete and thorough. They showed that
emerging particle picture is not point-like but takes the form of extended
quantum object which internal space-time structure is characterized by
Wigner's O(3)-little group. Another consequence resulting from Kim and Noz
solutions is that particle mass (or rather the mass of extended quantum
object) is generated by oscillatory-like field vibrations.

The purpose of this section is to describe simply quantum-mechanical
structure, called further the \textit{light-cone skeleton}, that underlies the
solutions of relativistic oscillator equation. In particular this
quantum-mechanical structure turns out to be very useful to show the way which
particle internal space-time structure becomes energy-dependent, as well as,
the way which particle motion can be expressed in terms of space-time
localized quantum states.

\subsection{Composite structure of massive state}

It was shown \cite{PK267} that four-momentum of relativistic particle with
mass can be made up of two light-cone momenta of massless particles. Since
each of the vectors of Minkowski space can be given in form of two superposed
light cone vectors, from the algebraic point of view, of course, there is no
surprise. The key point is, however, that in the case of on-mass-shell
four-momentum the magnitudes of these two light-cone momenta are related
through the scaling symmetry \cite{PK267}. This property is most easily
observed in $1+1$ dimensional gauge frame. Let us then recollect that
introduced in \cite{PK267} gauge frame was called the frame at which momentum
(or space) axis was chosen such to match the direction of particle motion. The
gauge frame description may then reduce the four-momentum to
\textit{bi-momentum }one.

So, let us consider a bimomentum $\binom{\Pi_{0}}{\Pi_{1}}$ given at the
\textit{photonic frame} (i.e. the frame which axes considered at the Minkowski
gauge frame are the light-cone axes) and assume that
\begin{equation}
\left(
\begin{array}
[c]{c}%
\Pi_{0}\\
\Pi_{1}%
\end{array}
\right)  =\left(
\begin{array}
[c]{cc}%
\frac{1}{\eta} & 0\\
0 & \eta
\end{array}
\right)  \binom{\frac{m_{0}c}{\sqrt{2}}}{-\frac{m_{0}c}{\sqrt{2}}},\label{bi1}%
\end{equation}
describes a relativistic particle with mass $m_{0}$. Indeed, at the Minkowski
gauge frame $\binom{\Pi_{0}}{\Pi_{1}}\rightarrow\binom{P_{0}}{P_{1}}$ where
\begin{equation}
\left(
\begin{array}
[c]{c}%
P_{0}\\
P_{1}%
\end{array}
\right)  =\left(
\begin{array}
[c]{cc}%
\frac{1}{\sqrt{2}} & \frac{-1}{\sqrt{2}}\\
\frac{1}{\sqrt{2}} & \frac{1}{\sqrt{2}}%
\end{array}
\right)  \left(
\begin{array}
[c]{c}%
\Pi_{0}\\
\Pi_{1}%
\end{array}
\right)  ,\label{bi2}%
\end{equation}
one finds that
\begin{equation}
\left(
\begin{array}
[c]{c}%
P_{0}\\
P_{1}%
\end{array}
\right)  =\left(
\begin{array}
[c]{c}%
\gamma m_{0}c\\
-\gamma\beta m_{0}c
\end{array}
\right)  ,\label{bi3}%
\end{equation}
where
\begin{equation}
\gamma=cosh\xi\text{ \ \ \ and \ \ \ }\gamma\cdot\beta=sinh\xi,\label{bi4}%
\end{equation}
for $\xi=ln\eta.$ Decomposition of Minkowski bi-momentum (\ref{bi3}) at the
photonic frame was shown on \textbf{Fig.1.
\begin{figure}
[ptb]
\begin{center}
\includegraphics[
height=2.7614in,
width=2.4785in
]%
{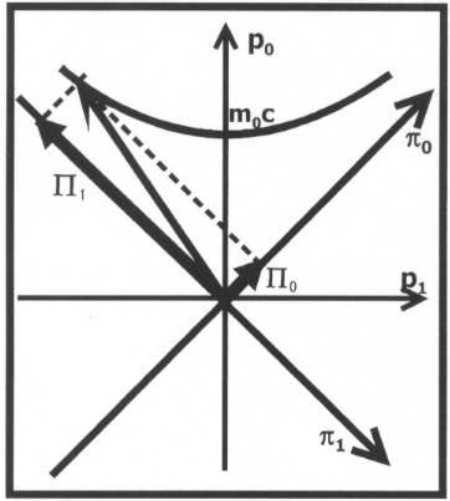}%
\caption{Decomposition of Minkowski four-momentum in the photonic frame. The
amplitudes of light-cone vectors $\Pi_{0}$ and $\Pi_{1}$ are related through
the scaling factor $\eta$, namely: $\Pi_{0}=\frac{1}{\eta}\frac{m_{0}c}%
{\sqrt{2}}$ and $\Pi_{1}=\eta\frac{m_{0}c}{\sqrt{2}}.$}%
\end{center}
\end{figure}
}The question one may ask, however, is whether the orthogonal transformation
(\ref{bi2}) is only an alternative of bi-momentum expression, or it reveals
also a composite structure of a massive state. Note, that following two
bi-momenta $\binom{\Pi_{0}}{0}$ and $\binom{0}{-\Pi_{0}}$ given at the
photonic frame, at the Minkowski gauge frame turn into $\binom{\Pi_{0}%
/\sqrt{2}}{\Pi_{0}/\sqrt{2}}$ and $\binom{\Pi_{0}/\sqrt{2}}{-\Pi_{0}/\sqrt{2}%
},$ which describe two photon states having opposite momenta, thereby opposite
propagation directions. A possibility of expression of massive state in terms
of two massless excitations propagating the opposite way rises, however,
another essential question, namely, how to explain that two such excitations
can describe a particle movement at all. The next two subsections are aimed to
approach to these issues in the most elementary way.

\subsection{Idea of \textit{light-cone skeleton}}

A natural consequence of the assumption about composite particle structure
must be departure from point-particle picture. It is widely believed that such
departure has been already done in terms of wave packet description, which is
basically true. Nevertheless, the point-particle picture still remains valid
as the classical one. Therefore, to emphasize the difference between
point-particle and extended object conceptual views we introduce the idea of
\textit{light-cone skeleton} intended to describe internal space-time
structure of particle with mass. As pointed out, the idea of light-cone
skeleton is purely quantum-mechanical one, thus, if correct, it should play a
similar role in field theory approach as the postulates of Planck and de
Broglie do in Schr\"{o}dinger theory.

In order to introduce extended quantum object description we start with
generic example of circle of radius $\lambda_{C}/2$ centered at the origin of
photonic (or Minkowski gauge) frame, as shown on Fig.2a\textbf{, }and assume
that this circle embraces a space-time area where extended quantum object is
localized. In terms of wave description one may expect this area to indicate
the maximum of probability density of quantum object distribution. Let us
assume also that the geometry of considering space-time region is the ground
state geometry, thereby the geometry of extended quantum object at rest (this
assumption is to become clear later).%
\begin{figure}
[ptb]
\begin{center}
\includegraphics[
height=2.418in,
width=4.1277in
]%
{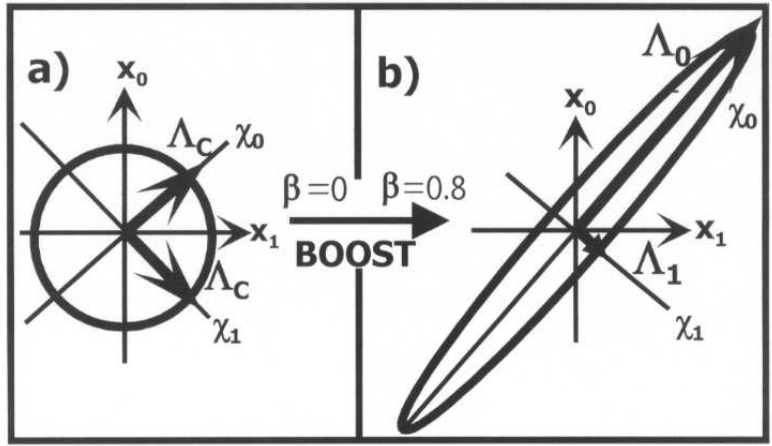}%
\caption{Boost deformation of extended quantum object. (a) The circle
represents space-time area of quantum object staying at rest. (b) At higher
(kinetic) energy level extended quantum object become localized at space-time
elliptical region. The vectors $\Lambda_{1}$ and $\Lambda_{2}$ were called the
light-cone skeleton. The boost transformation then corresponds to light-cone
skeleton deformation described by Eq. 8. Presented figure (almost) is
reprinted from [4].}%
\end{center}
\end{figure}

Next, let us consider another space-time area, now taking the form of an
ellipse (i.e. a deformed circle), as shown on Fig.2b\textbf{, }and assume that
it describes the same quantum object but already at a higher energy level. Let
as assume also that this higher energy level (established at the observer rest
frame) corresponds to an excited particle state that relates to particle
motion (i.e. the motion which is observed at the observer rest frame too). The
two light cone vectors $\Lambda_{0}$ and $\Lambda_{1},$ which, as it comes
from Fig.2b\textbf{, } ``spreads out'' the ellipse along its major and minor
axes, will be called further the \textit{light-cone skeleton.}

Although there is no need to assign any special value to $\lambda_{C}$\ , and
thus to $\Lambda_{0}$ and $\Lambda_{1}$, for illustrative purpose it is
advisable to consider a particle of \textit{model shape. }Model shape particle
will be called the particle which light-cone skeleton extensions $\lambda_{0}$
and $\lambda_{1}$, as depicted on Fig.3, are de Broglie wavelengths of
light-cone momenta (\ref{bi1}). The extensions of model shape particle then
are directly related to its dynamical features according to
\begin{equation}
\Pi_{0}=\frac{h}{\lambda_{0}}\text{ \ \ \ and \ \ }\Pi_{1}=-\frac{h}%
{\lambda_{1}}.\label{def2}%
\end{equation}
So, the light-cone skeleton for particle of model shape is
\begin{equation}
\left(
\begin{array}
[c]{c}%
\Lambda_{0}\\
\Lambda_{1}%
\end{array}
\right)  =\left(
\begin{array}
[c]{c}%
\frac{1}{2}\lambda_{0}\\
\frac{1}{2}\lambda_{1}%
\end{array}
\right)  =\left(
\begin{array}
[c]{c}%
\frac{1}{2}\frac{1}{\sqrt{2}}\eta\lambda_{C}\\
\frac{1}{2}\frac{1}{\sqrt{2}}\frac{1}{\eta}\lambda_{C}%
\end{array}
\right)  ,\label{def3}%
\end{equation}
where, due to (\ref{bi1})
\begin{equation}
\lambda_{C}=\frac{h}{m_{0}c},\label{Cw}%
\end{equation}
is the Compton wavelength.%
\begin{figure}
[ptb]
\begin{center}
\includegraphics[
height=2.5581in,
width=2.6455in
]%
{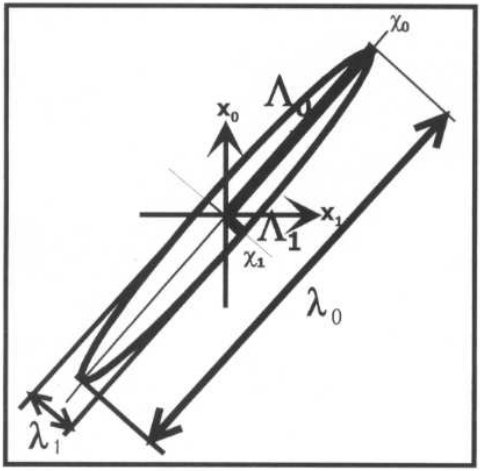}%
\caption{The particle of ``model shape''. The light-cone skeleton extensions
of particle of model shape, $\lambda_{0}$ and $\lambda_{1}$, correspond
exactly to its energy and momentum, see formulas (2), (3) and (5).}%
\end{center}
\end{figure}

The light-cone skeleton structure then links particle dynamical and
geometrical properties. Indeed, one easily finds that there is $1:1$
correspondence between particle dynamical features expressed in terms of
light-cone momenta (\ref{bi1}) and $circle\rightarrow ellipse$ transition
described by light-cone skeleton deformation
\begin{equation}
\left(
\begin{array}
[c]{c}%
\Lambda_{C}\\
\Lambda_{C}%
\end{array}
\right)  \rightarrow\left(
\begin{array}
[c]{c}%
\Lambda_{0}\\
\Lambda_{1}%
\end{array}
\right)  =\left(
\begin{array}
[c]{c}%
\eta\Lambda_{C}\\
\frac{1}{\eta}\Lambda_{C}%
\end{array}
\right)  ,\label{def1}%
\end{equation}
where, due to (\ref{def3}), $\Lambda_{C}=\lambda_{C}/2\sqrt{2}$ (cf. Fig.2).
So, one finds that along with particle (kinetic) energy increase the area of
its space-time localization becomes more and more elongated \cite{Kim98}. Such
behavior, of course, is unpredictable within the framework of Schr\"{o}dinger
quantum mechanics. On the other hand, it becomes visible that special
relativity put in ``exclusive hands of quantum mechanics'' must provide a
description of particle motion quite different from that, which by making use
of classical approach, we get used to it.

\subsection{Moving particle and space-time localized quantum states}

Particle kinematics expressed in terms of space-time localized quantum states
naturally blends into the landscape of quantum mechanics based on the ground
of the covariant and preferred frame description. In such environment
difficulties related to time and length measure do not even have a chance to
appear, so that the relativity issue becomes ripped its ``mystery cover'' off.
One of the consequence resulting from that is that relativity aspect is fund
to provide a number interesting insights about internal particle space-time
structure. Given that we do have a relativistic wave equation that yield
space-time localized solutions, let us try to draw some basic conclusions
resulting from that.

To apply the earlier results let us consider a particle of \textit{model
shape} which ground state probability distribution is centered in circle of
radius $\lambda_{C}/2.$ In such case, as noticed, the particle must stay at
rest, so that its Minkowski frame bi-momentum $\binom{P_{0}}{P_{1}}%
=\binom{m_{0}c}{0}.$ Along with particle energy increase its space-time
structure is supposed to become changed as well. Indeed, the boost
transformation that induces $\binom{m_{0}c}{0}\rightarrow\binom{\gamma m_{0}%
c}{\gamma\beta m_{0}c},$ must affect also the light-cone skeleton structure,
thereby the geometry of the state. Due to (\ref{def1}), the circle area of
ground state must be transformed into elliptical region of excited state. Let
us then explain the way which light-cone skeleton deformation shown on
Fig.2\textbf{ }provide informations about particle kinematics.

For that purpose, let us assume that effectively measured quantities at the
real space are not the light-cone ones but, similarly like in the case of
bi-momentum (\ref{bi2}), are the corresponding to them Minkowski
``equivalents''. Thus, the relationship between the Minkowski intervals and
space-time separations of quantum object expressed via light-cone skeleton
structure, as shown on Fig.4a\textbf{,} are
\begin{equation}
\left(
\begin{array}
[c]{c}%
\frac{\Delta x_{0}}{2}\\
\frac{\Delta x_{1}}{2}%
\end{array}
\right)  =\left(
\begin{array}
[c]{cc}%
\frac{1}{\sqrt{2}} & \frac{1}{\sqrt{2}}\\
-\frac{1}{\sqrt{2}} & \frac{1}{\sqrt{2}}%
\end{array}
\right)  \left(
\begin{array}
[c]{c}%
\Lambda_{1}\\
\Lambda_{2}%
\end{array}
\right)  .\label{lc1}%
\end{equation}
The physical meaning of intervals $\Delta x_{0}$ and $\Delta x_{1}$ seems to
be complementary. Indeed, by making use of textbook formulas which allow us to
combine scaling parameter $\eta$ (see (\ref{bi4})) with a velocity $w$
according to
\begin{equation}
\gamma=\frac{1}{\sqrt{1-w^{2}/c^{2}}}\text{ \ and \ }\beta=\frac{w}%
{c},\label{wvelo}%
\end{equation}
one obtains
\begin{equation}
\Delta x_{0}=\frac{1}{\sqrt{1-w^{2}/c^{2}}}\lambda_{C},\label{uc1}%
\end{equation}
and
\begin{equation}
\Delta x_{1}=\frac{w/c}{\sqrt{1-w^{2}/c^{2}}}\lambda_{C}\text{\ .}\label{uc2}%
\end{equation}
Note, that $\Delta x_{0}\geqslant\Delta x_{1}.$ In particular, for $w$
$\lessapprox$ $c,$ the both intervals are almost equal, but in the case of
$w=0$, $\Delta x_{0}=\lambda_{C}$ whereas $\Delta x_{1}=0.$ This suggests that
$\Delta x_{1}$ may be understood as the uncertainty of particle center (of
mass) position inside some ``quantum region'', in time period resulting just
from (\ref{uc1}). Thus, one may say that uncertainty $\Delta x_{1}$ applies to
\textit{point-particle }position indeed. Assuming that this
\textit{point-particle} moves (or oscillates) inside the mentioned quantum
area, one may call such movement a movement in a \textit{classical channel.}
On the other hand, time separation $\Delta x_{0}$ might be identified with the
time of life of temporarily localized quantum state (do not mistake with
particle life time), which classical channel\textit{\ }width (or uncertainty)
is just $\Delta x_{1}.$ Thus, the time separation interval $\Delta x_{0}$
might be called the uncertainty of \textit{quantum channel.%
\begin{figure}
[ptb]
\begin{center}
\includegraphics[
height=2.5322in,
width=3.6426in
]%
{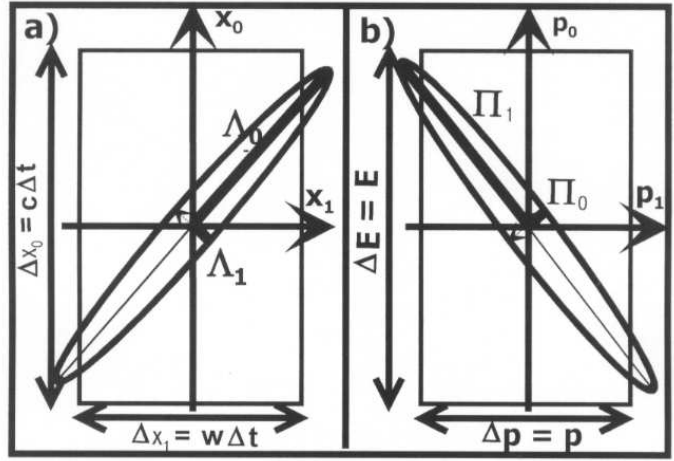}%
\caption{Two mutually conjugated light-cone structures given respectively in
position and momentum space. (a) The space-time area ``spread out'' on
light-cone skeleton structure in position space indicates maximum probability
density of quantum object distribution. One the other hand the very light cone
skeleton extensions projected on Minkowski frame axes provide informations
about particle kinematics (see formula (9)). Similarly, the light-cone
skeleton structure given in momentum space (b) determines energy-momentum
distribution of extended quantum object [4]. For the particle of model shape
its energy and momentum fluctuations amount to its energy and momentum
themselves (see formula (58)).}%
\end{center}
\end{figure}
}

To find the explicit dependence between the life times of quantum state of
moving particle and particle at rest, let as put
\begin{equation}
\lambda_{C}=c\Delta\tau\text{ \ \ and \ \ }\Delta x_{0}=c\Delta t,\label{uc10}%
\end{equation}
which means that intervals $\Delta\tau$ and $\Delta t$ are respectively the
time separations of ground state and excited state of particle motion. Then,
by combining formula (\ref{uc10}) with (\ref{uc1}) and (\ref{uc2}) one obtains
the following expressions
\begin{equation}
\Delta t=\frac{\Delta\tau}{\sqrt{1-w^{2}/c^{2}}},\label{uc11}%
\end{equation}
and
\begin{equation}
\Delta x_{1}=w\Delta t,\label{uc12}%
\end{equation}
which explain, so-called, time dilatation effect, now in pure
quantum-mechanical manner. It is clear that currently formula (\ref{uc11}) has
noting to do with ``relativistic time measure''. Instead, one finds that the
task of this ``relativistic effect'' is taken over by energy-dependent
deformation of particle internal space-time structure.

Description of particle motion based on temporarily localized quantum states
distinguishes the following two situations: the first one is when particle
life-time and the life-time of temporarily localized quantum state are the
same, and the second one is when particle life-time is substantially longer.
So, in the latter case the quantum picture of particle motion cannot be given
in a form of single space-time localized region, as shown on Fig.4a, but it
must consist of many such regions, as shown on Fig.5\textbf{.} The particle
motion then cannot be considered smooth, but rather as a jump-like or
\textit{oscillatory motion}. The analysis of covariant harmonic oscillator
solutions will bring us to this point again later on.%
\begin{figure}
[ptb]
\begin{center}
\includegraphics[
height=2.7423in,
width=2.1326in
]%
{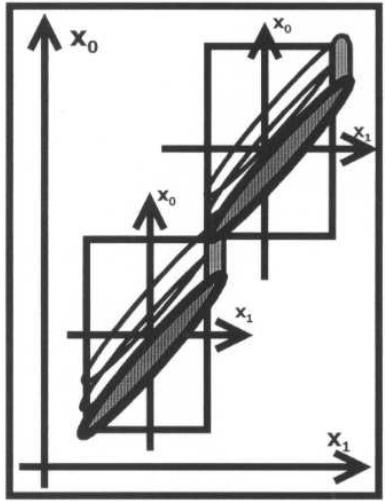}%
\caption{Oscillatory motion of extended quantum object. The motion induced by
kinetic and potential excitations (described in Section 5) effectively takes
the form of space-time probability current flow.}%
\end{center}
\end{figure}

And finally, let us note that formulas (\ref{uc1}), (\ref{uc2}) and (\ref{Cw})
enable us to express particle space-time extensions in form of Lorentz
invariant interval
\begin{equation}
\left(  \Delta x_{0}\right)  ^{2}-\left(  \Delta x_{1}\right)  ^{2}=\left(
\frac{h}{m_{0}c}\right)  ^{2}.\label{inv1}%
\end{equation}
which, in clear-cut way, points out to the relationship between the special
relativity and quantum mechanics.

\section{Wave description of extended quantum object}

Relativistic oscillator wave function has been considered already by Yukawa in
1953 \cite{Yu53} who attempted to model the behavior of relativistic particles
having internal structures. Later this issue also was undertaken by Feynman
\cite{Fosc} who paradoxically suggested himself to use relativistic oscillator
wave functions instead of \textit{his }diagrams\textit{\ }for study hadronic
structures and interactions\textit{. }Nevertheless, the powerful method of
Feynman diagrams has effectively overshadowed many interesting results of
bound states analysis carried out strictly within the boundaries of covariant approach.

The main purpose of this section is to show that Feynman equation equipped
with Kim and Noz solutions provides two seemingly quite different particle
images. The first point-particle image is associated with running waves (and
thus with underlaying structure of Feynman diagrams). The other particle image
emerges from normalizable wave functions and takes the form of extended
quantum object.

\subsection{Feynman equation and Kim and Noz solutions}

The differential equation of Feynman, Kislinger and Ravndal \cite{Fosc},
namely
\begin{equation}
\left\{  \left[  \left(  \frac{\partial}{\partial x_{a}^{\mu}}\right)
^{2}+\left(  \frac{\partial}{\partial x_{b}^{\mu}}\right)  ^{2}\right]
-\frac{\kappa}{8}\left(  x_{a}^{\mu}-x_{b}^{\mu}\right)  ^{2}\right\}
\phi(x_{a,}x_{b})=0,\label{F1}%
\end{equation}
was aimed to describe a hadron consisting of two quarks at $x_{a}$ and
$x_{b},$ bound together by a harmonic oscillator potential of strength
$\kappa$. Actually, Eq. (\ref{F1}) is the Kim and Noz version of Feynman
equation \cite{Kbook}, where, additionally, the potential constant $\kappa,$
now, is kept in the explicit form to emphasize the meaning of units of
$\kappa=1$ later on. Furthermore, since the algebraic structure of quarks is
not taken into account, it is enough to assume that $x_{a}$ and $x_{b}$ are
the coordinates of self-interacting scalar quantum object.

Following the procedure of Kim and Noz let us introduce the new variables:
\begin{equation}
X=\frac{x_{a}+x_{b}}{2},\label{X}%
\end{equation}
which specifies the quantum object space-time position and
\begin{equation}
x=\frac{x_{a}-x_{b}}{2},\label{x}%
\end{equation}
which determines quantum object space-time separations. The new variables make
possible to write down Eq. (\ref{F1}) in explicitly separable form
\begin{equation}
\left\{  \left(  \frac{\partial}{\partial X^{\mu}}\right)  ^{2}+\left[
\left(  \frac{\partial}{\partial x^{\mu}}\right)  ^{2}-\kappa\left(  x^{\mu
}\right)  ^{2}\right]  \right\}  \phi(X,x)=0.\label{F2}%
\end{equation}
Indeed for
\begin{equation}
\phi(X,x)=f(X)\cdot\psi(x),\label{sol1}%
\end{equation}
one finds that following two differential equations must be satisfied:
\begin{equation}
\left\{  \left(  \frac{\partial}{\partial X^{\mu}}\right)  ^{2}+\left(
\frac{Mc}{h}\right)  ^{2}\right\}  f(X)=0,\label{KG}%
\end{equation}
and
\begin{equation}
\left[  \left(  \frac{\partial}{\partial x^{\mu}}\right)  ^{2}-\kappa\left(
x^{\mu}\right)  ^{2}\right]  \psi(x)=\left(  \frac{Mc}{h}\right)  ^{2}%
\psi(x).\label{cov1}%
\end{equation}
Equation (\ref{KG})\ is a Klein-Gordon equation for particle with mass $M$.
The solutions of this equation are the plane waves
\begin{equation}
\ f(X)=f_{0}e^{\pm\left(  P\nu X^{\nu}\right)  },\label{pl}%
\end{equation}
which, indeed, call the image of point-particle carrying out the four-momentum
$P_{\nu},$ where $P_{\nu}P^{\nu}=M^{2}c^{2}.$ Currently, however, relativistic
particle is described also by Eq. (\ref{cov1}). But Eq. (\ref{cov1})
determines not only the value of particle mass, but also it dresses it up with
internal space-time structure. To observe this, as well as, because of the
fact that solutions of (\ref{cov1}) are naturally given in dimensionless
units, it is advisable to write down Eqs. (\ref{KG}) and (\ref{cov1}) in
reduced form. For that purpose let us introduce some characteristic length
scale $\lambda_{c}/2$ and rewrite Eq. (\ref{KG}) as
\begin{equation}
\left\{  \left(  \frac{\partial}{\partial\widetilde{X}^{\nu}}\right)  ^{2}%
+\mu^{2}\right\}  f(\widetilde{X})=0,\label{KG2}%
\end{equation}
where $\widetilde{X}^{\nu}=\frac{X^{\nu}}{\lambda_{c}/2},$
\begin{equation}
f(\widetilde{X})=f_{0}e^{\pm\left(  \widetilde{P}\nu\widetilde{X}^{\nu
}\right)  },\label{pl2}%
\end{equation}
so that
\begin{equation}
\widetilde{P}\nu\widetilde{P}^{\nu}=\mu^{2}.\label{L1}%
\end{equation}
Similarly, in the case of Eq. (\ref{cov1}) one obtains
\begin{equation}
\left[  \left(  \frac{\partial}{\partial\widetilde{x}^{\mu}}\right)
^{2}-\kappa\left(  \frac{\lambda_{c}}{2}\right)  ^{4}\left(  \widetilde
{x}^{\mu}\right)  ^{2}\right]  \psi(\widetilde{x})=\mu^{2}\psi(\widetilde
{x}),\label{cov2}%
\end{equation}
where $\widetilde{x}^{\mu}=\frac{x^{\mu}}{\lambda_{c}/2}.$ Additionally, if
one puts
\begin{equation}
\kappa\left(  \frac{\lambda_{c}}{2}\right)  ^{4}=1,\label{k1}%
\end{equation}
one finds that notation that involves unit potential strength makes use of
intrinsically built in length scale $\lambda_{c}$. Elementary solution of Eq.
(\ref{F2}) is then a plane wave $f(\widetilde{X})$ which amplitude is
localized in some space-time region, if only $\psi(\widetilde{x})$ is entirely
normalizable function.

To find the solutions of Eq. (\ref{cov2}) let us assume that condition
(\ref{k1}) is fulfilled and (by skipping the tildes from now on) let us
rewrite this equation in more explicit form
\begin{equation}
\left[  \left(  \frac{\partial^{2}}{\partial x_{0}^{2}}-\frac{\partial^{2}%
}{\partial\mathbf{x}^{2}}\right)  +\left(  x_{0}^{2}-\mathbf{x}^{2}\right)
\right]  \psi_{\lambda}(x)=\mu^{2}\psi_{\lambda}(x).\label{cov3}%
\end{equation}
Eq. (\ref{cov3}) was just the starting point of Kim and Noz analysis
\cite{Kbook}. They considered next the set of following solutions
\begin{align}
\psi_{\lambda}(x)  &  =\left(  \frac{1}{\pi^{1/4}}\right)  ^{4}\left(
\frac{1}{2}\right)  ^{(k+l+m+n)/2}\left(  \frac{1}{k!l!m!n!}\right)
^{1/2}\label{psi1}\\
&  \times H_{k}(x_{0})H_{l}(x_{1})H_{m}(x_{2})H_{n}(x_{3})\nonumber\\
&  \times exp\left(  -\frac{1}{2}\left(  x_{0}^{2}+x_{1}^{2}+x_{2}^{2}%
+x_{3}^{2}\right)  \right)  ,\nonumber
\end{align}
where $H_{k}(x_{0}),$ $H_{l}(x_{1})...$ are the Hermite polynomials and the
eigenvalue
\begin{equation}
\mu^{2}=1+l+m+n-k\equiv1+\lambda,\label{L2}%
\end{equation}
so that $\lambda=l+m+n+k$ where $k,l,m$ and $n$ are integer numbers. So,
indeed the wave functions $\psi_{\lambda}(x)$ are entirely normalizable,
however, the covariant oscillator problem turns out to be infinitely
degenerated. To limit the degeneracy to a finite number, as well as, to retain
only those solutions endowed with transparent physical meaning, Kim and Noz
imposed a covariant condition that suppressed the time-like oscillations,
thereby avoided the problem of negative energies too. The wave function
(\ref{psi1}) then has been reduced to
\begin{align}
\psi_{\lambda^{\prime}}(x)  &  =\left(  \frac{1}{\pi^{1/4}}\right)
^{4}\left(  \frac{1}{2}\right)  ^{(l+m+n)/2}\left(  \frac{1}{l!m!n!}\right)
^{1/2}\label{psi2}\\
&  \times H_{l}(x_{1})H_{m}(x_{2})H_{n}(x_{3})\nonumber\\
&  \times exp\left(  -\frac{1}{2}\left(  x_{0}^{2}+x_{1}^{2}+x_{2}^{2}%
+x_{3}^{2}\right)  \right)  ,\nonumber
\end{align}
where
\begin{equation}
\lambda^{\prime}=l+m+n.\label{L3}%
\end{equation}
Kim and Noz showed that wave functions (\ref{psi1}) constitute irreducible,
infinite-dimensional unitary representation of the Lorentz group which
describes \textit{Klein-Gordon-like} particles with definite momentum $P_{\mu
}$ and internal space-time structure but devoid of time-like oscillations. We
will make use of this crucial property later on. First, however, it is
advisable to extract a physical content of solutions (\ref{pl2}) and
(\ref{psi2}) and correlate the ``shape'' of particle internal structure with
its four-momentum $\widetilde{P}_{\mu}$ .

\subsection{Geometry of the ground state}

Kim and Noz showed that wave functions (\ref{psi1}) form a vector space for
O(3)-like little group. However, the symmetry of very ground state
($l=m=n=0$),
\begin{equation}
\psi_{0}(x)=\frac{1}{\pi}exp\left(  -\frac{1}{2}\left(  x_{0}^{2}+x_{1}%
^{2}+x_{2}^{2}+x_{3}^{2}\right)  \right)  ,\label{g1}%
\end{equation}
turns out even to be the symmetry of four-dimensional sphere. Keeping in mind
that function (\ref{g1}) is given in $\lambda_{c}/2$ units, one finds that the
radius of ``ground state sphere'' is $1/\sqrt{2}$ . Indeed, since the ground
state function (\ref{g1}) separates according to
\begin{equation}
\psi_{0}(x)=\frac{e^{-\frac{\chi_{0}^{2}}{2}}}{\pi^{1/4}}\frac{e^{-\frac
{\chi_{1}^{2}}{2}}}{\pi^{1/4}}\frac{e^{-\frac{\chi_{2}^{2}}{2}}}{\pi^{1/4}%
}\frac{e^{-\frac{\chi_{3}^{2}}{2}}}{\pi^{1/4}},\label{g2}%
\end{equation}
the same concerns the probability density distribution
\begin{equation}
\left|  \psi_{0}(x)\right|  ^{2}=\frac{e^{-x_{0}^{2}}}{\pi^{1/2}}%
\frac{e^{-x_{1}^{2}}}{\pi^{1/2}}\frac{e^{-x_{2}^{2}}}{\pi^{1/2}}%
\frac{e^{-x_{3}^{2}}}{\pi^{1/2}},\label{gg2}%
\end{equation}
where each of the factors appear to be the Gaussian distributions of the same
dispersion $2\sigma=\sqrt{2}$. Thus, if one intersect the sphere $x_{0}%
^{2}+x_{1}^{2}+x_{2}^{2}+x_{3}^{2}=\frac{1}{2}$ with e.g. $x_{0}x_{1}$ plane
one finds that included at this plane light-cone skeleton of coordinates
$\binom{1/\sqrt{2}}{1/\sqrt{2}}$ (cf. Fig.2a) becomes representative for the
whole quantum object described by Eq. (\ref{g2}). Indeed, accordingly to
(\ref{lc1}) the light-cone skeleton structure with equal ``arms'' must
describe quantum object at rest. Note also that the form of wave function
(\ref{g2}) is invariant with respect to transformation
\begin{equation}
\left(
\begin{array}
[c]{c}%
\chi_{0}\\
\chi_{1}%
\end{array}
\right)  =\left(
\begin{array}
[c]{cc}%
\frac{1}{\sqrt{2}} & \frac{1}{\sqrt{2}}\\
\frac{-1}{\sqrt{2}} & \frac{1}{\sqrt{2}}%
\end{array}
\right)  \left(
\begin{array}
[c]{c}%
x_{0}\\
x_{1}%
\end{array}
\right)  ,
\end{equation}
which replaces the coordinates of Minkowski frame $x_{0},x_{1},$ with those of
photonic (or light-cone) ones $\chi_{0},\chi_{1}$. Indeed, since
\begin{equation}
\chi_{0}^{2}+\chi_{1}^{2}=x_{0}^{2}+x_{1}^{2},
\end{equation}
alternative form of wave function (\ref{g2}) is
\begin{equation}
\psi_{0}(x,\chi)=\frac{e^{-\frac{\chi_{0}^{2}}{2}}}{\pi^{1/4}}\frac
{e^{-\frac{\chi_{1}^{2}}{2}}}{\pi^{1/4}}\frac{e^{-\frac{x_{2}^{2}}{2}}}%
{\pi^{1/4}}\frac{e^{-\frac{x_{3}^{2}}{2}}}{\pi^{1/4}}.\label{g3}%
\end{equation}
High symmetry of the ground state makes then that distinction between the
coordinates of Minkowski and photonic frames is covered.

In the next section we will show that images of point-particle and
extended-particle outlined above are indeed complementary. The light-cone
skeleton concept will make us possible to superimpose the both particle
pictures. Furthermore, we will indicate the way which particle motion can be
described in terms of space-time localized wave packets. So, the issue of
normalizable space-time solutions becomes of special interest especially in
the context of currently lunching idea that special relativity and quantum
mechanic make an indivisible whole.

\section{Localized states of motion}

Currently discussed approach making the special relativity basically a quantum
mechanics toll, enormously simplifies the whole ``relativity'' issue. Lorentz
symmetry becomes no longer identified with the relative motion of inertial
frames, thereby the relativity aspect of length and time measure becomes
completely withdrawn from the relativity issue along with its orthodox mode.
Instead, Lorentz symmetry group turns out to tackle the problem of description
of energy-dependent geometry of extended quantum objects.

It was shown \cite{PK267} that quantum-mechanical symmetry related to the
freedom of choice of comparison scale combined with Euclidean rotations led to
the concept of Lorentz group. The covariant form of differential equations
encompasses then the symmetries of scaling and $3d$ rotations. From the
physical point of view, however, more important seem to be the conclusions
resulting form solutions of these equations. Given that one really has the
solutions that describe quantum object internal space-time structure, the
Lorentz group then is expected to disclose many interesting particle features.
In this section we make use of Kim and Noz solutions to indicate the way which
space-time localized quantum states tackle description of particle motion. The
analysis corresponds well to this given in section 2.3, but now, of course,
goes much beyond the pure quantum-mechanical considerations.

\subsection{\textit{Kinetic} excitations}

Basing on the results of preceding section one finds that complete wave
function of the ground state of Eq. (\ref{F2}) is
\begin{equation}
\phi_{0}(X,x)=e^{\pm m_{0}cX^{0}}\cdot\frac{1}{\pi}e^{-\frac{1}{2}\left(
x_{0}^{2}+x_{1}^{2}+x_{2}^{2}+x_{3}^{2}\right)  }.\label{gsf1}%
\end{equation}
(Note that energy $E=m_{0}c^{2}=hc/\lambda_{C}$ corresponds to $\mu=1,$ i.e.
the energy expressed in $\lambda_{C}/2$ units). The wave function (\ref{gsf1})
describes then a relativistic scalar at rest.

The form of wave function (\ref{sol1}), or (\ref{gsf1}), suggests that
solutions of relativistic oscillator equation involves two kinds of
excitations. The first kind is related to, let say, \textit{potential
excitations, }i.e\textit{. }excitations induced by vibrations of particle
internal structure. According to (\ref{L1}) and (\ref{L2}) one finds that
potential excitations are to be characterized by different mass levels.

Another kind of excitations, which, as it comes also from analysis before, is
expected to be related directly to particle kinematics. Let us call this kind
of excitations the \textit{kinetic excitations. }In fact the very approach
based on preferred frame description invokes the space-time localized quantum
states to be the states of particle motion. Let us then determine the notion
of \ ``kinetic excitations'' in more precise manner.

The simplest and most natural way to obtain the state of particle motion is to
boost the ground state (or any other state being assumed to describe particle
at rest). Let us then apply a Lorentz boost to the ground state (\ref{gsf1}).
The running wave factor must be transformed according to
\begin{equation}
e^{\pm\left(  m_{0}cX^{0}\right)  }\rightarrow e^{\pm\left(  P^{\prime}\nu
X^{\prime\nu}\right)  },\label{b1}%
\end{equation}
where the components of $P_{\nu}=(P_{0},P_{1},0,0)$ are those given in
(\ref{bi3}). On the other hand, due to (\ref{g3}), the transformation of
standing wave factor must be
\begin{equation}
e^{-\frac{1}{2}\left(  \chi_{0}^{2}+\chi_{1}^{2}+x_{2}^{2}+x_{3}^{2}\right)
}\rightarrow e^{-\frac{1}{2}\left(  \left(  \eta\chi_{0}\right)  ^{2}+\left(
\frac{1}{\eta}\chi_{1}\right)  ^{2}+x_{2}^{2}+x_{3}^{2}\right)  }.\label{b2}%
\end{equation}
So, the boost action taken along the gauge $x_{1}$ (or $\chi_{1}$) direction
upon the ground state (\ref{gsf1}) yields the wave faction
\begin{equation}
\phi_{0,\eta}(X,x,\chi)=e^{\pm P_{\mu}X^{\mu}}\cdot\frac{1}{\pi}e^{-\frac
{1}{2}\left(  \left(  \eta\chi_{0}\right)  ^{2}+\left(  \frac{1}{\eta}\chi
_{1}\right)  ^{2}+x_{2}^{2}+x_{3}^{2}\right)  }.\label{gsf2}%
\end{equation}
The coordinates $X$ and $x$ are, of course independent ones, nevertheless,
from the physical point of view, they must refer to the same area of the
observer rest frame. Thus, the wave function (\ref{gsf2}) effectively assumes
a form of a plane wave running inside some ellipsoidal cover, as shown on Fig
6a.\textbf{\ }Thereby the momentum transfer $P_{\mu}$ carried out by this
plane wave is to be localized inside that ellipsoidal cover too. This kind of
four-momentum transfers, i.e. the transfers induced by space-time localized
running waves have been already called the \textit{kinetic excitations}.%
\begin{figure}
[ptb]
\begin{center}
\includegraphics[
height=2.0505in,
width=3.4662in
]%
{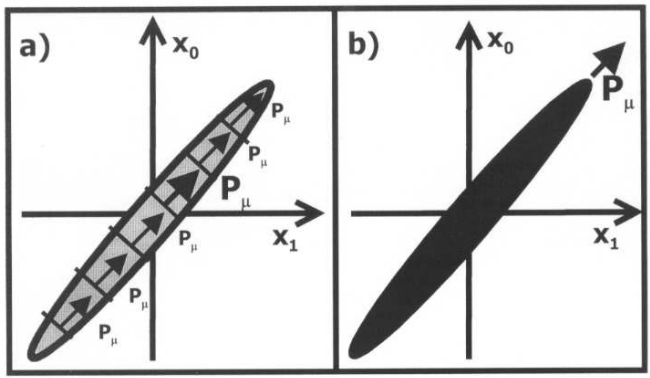}%
\caption{Kinetic and potential excitations - a new image of particle wave
duality. (a) The running plane waves carrying out the four-momentum $P_{\mu}$
and included in space-time localized region (being ``spread out'' on
light-cone skeleton structure $\Lambda_{0},\Lambda_{1}$) are called the
kinetic excitations. On the other hand the same four-momentum $P_{\mu}$ can be
carried out by standing waves embraced by the same space-time localized region
(b). This kind of standing waves is called the potential excitations. The
potential excitations then represents compact quantum object running through
the space.}%
\end{center}
\end{figure}

There are two important remarks that help to understand better the case of
kinetic excitations. The first is that action of Lorentz boost considered as
the transformation of Minkowski frame is to be regraded as the passive
transformation. In other words, the Lorentz (or Poincar\'{e}) symmetry group
appears the symmetry of quantum states themselves. But this, in turn, means
that time and length measures of all inertial frame are always the same. It
was noticed \cite{PK267} that relativistic description requires a clear
distinction between the \textit{vital} and \textit{frozen} time meanings. To
recollect, it was explained that vital time is the measure of pace of observed
changes and does not undergo relativistic transformation rules. In contrary to
this the frozen time, which gives us energy measure in sense of inverse time
units, does. Above analysis then suggests that frozen time meaning enters also
description of quantum object extensions.

The second thing is that probability distributions provided by normalizable
solutions of covariant oscillator equation are Lorentz invariants
\cite{KimPR87},\cite{KimPR88}. In particular the space-time area of maximum
probability density for kinetic excitation (\ref{gsf2}), due to light-cone
skeleton deformation (\ref{def1}), is proportional to $\lambda_{1}%
\times\lambda_{2}$ $\sim1/m_{0}^{2}.$ These features, however, become visible
only in the light-cone coordinate frame. It was Kim who suggested to use the
light-cone coordinate system as a natural language for Lorentz covariant
phase-space representation of quantum mechanics \cite{KimPR87}. It seems,
however, that even more likely is a scenario at which descriptions based
simultaneously on the Minkowski, as well as, light-cone frame, play the roles
that are of equal importance and complementary. The issue, one may say,
resembles almost a watching of $3d$ object at the real space. To get its
complete image, in general, one needs to look at it more than just only one of
its side. Similarly, a picture of extended quantum object emerging from the
solutions of relativistic oscillator equation needs the light-cone coordinate
system to become visible. On the other hand, the other quantum object
properties such as its four-momentum or the time of life need the Minkowski
frame (observation) to be established.

\subsection{\textit{Potential} excitations}

The potential excitations were identified with the vibrations of internal
particle structure, thereby describing them standing waves (\ref{psi2}) might
be called the \textit{potential }states too. Similarly like in the case of
kinetic excitations, the issue of potential states considered within the
context of preferred frame description gives rise to a question about their
physical meaning. Especially, what happens if the potential state is boosted
and what is the influence on its physical meaning then? In this subsection we
focus mainly on the first part of this question.

Generic example of the boost of the ground state (\ref{b2})\ showed us that
along with particle energy increase the space-time area embracing localized
probability distribution, by simultaneous elongation and shrinking along the
two orthogonal light-cone directions, undergoes the deformations too. The use
photonic frame then enables potential states to become useful probabilistic
toll against the background of relativistic preferred frame description. The
important question is, however, whether the boost action affects the initial
character of the excitations mode. In particular, whether it provides an
admixture of ``something else'' that goes beyond the pure potential
excitations. Fortunately the analysis of Kim and Noz in clear-cut way has
resolved this problem.

Kim and Noz have shown that wave functions (\ref{psi1})\ constitute a linear
infinite-dimensional unitary representation of the Lorentz group. As a result,
one finds that potential state $\psi_{\lambda^{\prime}}(x),$ if boosted along
$x_{1}$ direction, turns into the boosted one $\psi_{\lambda^{\prime},\eta
}(x),$ which, in turn, can be made up of unboosted potential states again,
namely
\begin{align}
\psi_{\lambda^{\prime},\eta}(x)  &  =\frac{1}{\pi}\left(  \frac{1}{2}\right)
^{\frac{l+m+n}{2}}\left(  \frac{1}{l!m!n!}\right)  ^{\frac{1}{2}}H_{m}%
(x_{2})H_{n}(x_{3})\label{psi3}\\
&  \times\left(  \frac{1}{2}\left(  \eta+\frac{1}{\eta}\right)  \right)
^{-(1+l)}exp\left(  -\frac{1}{2}\left(  x_{0}^{2}+x_{1}^{2}\right)  \right)
\nonumber\\
&  \times\left(  \underset{k=0}{\overset{\infty}{\sum}}\left(  \frac{1}%
{4}\right)  ^{k}\left(  \frac{\eta-\frac{1}{\eta}}{\eta+\frac{1}{\eta}%
}\right)  ^{k}H_{k}(x_{0})H_{l+k}(x_{1})\right)  ,\nonumber
\end{align}
So, the boost transformation performed on any potential state does not change
its potential character at all. One needs to emphasize, however, that
time-like excitations that do not appear at the level of effectively observed
on-mass-shell, become inevitable components of internal (hidden-like) particle
structure. It is worthwhile to notice also that notion of ``particle rest
frame'' (in contrary to ``observer rest frame'') turns out to be very
tantalizing indeed.

The remaining aspect of the physical meaning of potential state boost is to be
discussed next.

\section{A new particle-wave duality picture emerging form relativistic
oscillator model}

The idea of particle-wave duality undoubtedly is one of the most crucial
physical ideas that moulds our physical intuitions. The postulate of de
Broglie
\begin{equation}
p=\frac{h}{\lambda},\label{du1}%
\end{equation}
which combines particle momentum $p$ with the wavelength $\lambda,$ and the
postulate of Planck
\begin{equation}
E=\frac{h}{T},\label{du2}%
\end{equation}
which relates particle energy $E$ to the wave period $T$, set up the
quantum-mechanical basis for this idea. The Schr\"{o}dinger equation that put
the postulates of Planck and de Broglie in ``true'' wave description reality,
gave us a particle-wave duality picture, which, at the highest simplicity, is
the following: the free particles are point-like whereas their wave-like
nature is (successfully) represented by running (plane) waves or plane-wave
packets. The key point is, however, that relativistic quantum mechanics cannot
not improve this picture essentially, unless the predictions resulting from
the solutions of covariant harmonic oscillator equation become appreciated.
The aim of this section is to provide a new picture of particle-wave duality
emerging just form the relativistic oscillator equation. Although this new
picture goes much beyond the framework of Schr\"{o}dinger approach, the old
Schr\"{o}dinger painting seems to be encompassed by the new one rather than challenged.

\subsection{Oscillatory motion of extended quantum object}

The explicit form of wave function (\ref{gsf2}) written at the Minkowski
frame, due to (\ref{psi3}), is
\begin{equation}
\phi_{0,\eta}(X,x)=e^{\pm\frac{i}{\hbar}\left(  P_{\mu}X^{\mu}\right)  }%
\cdot\left(  \frac{1}{\pi}e^{-\frac{1}{2}\left(  x_{0}^{2}+x_{0}^{2}+x_{2}%
^{2}+x_{3}^{2}\right)  }\Psi_{\eta}(x_{0},x_{1})\right)  ,\label{fi1}%
\end{equation}
where
\begin{equation}
\Psi_{\eta}(x_{0},x_{1})=\left(  \frac{1}{2}\left(  \eta+\frac{1}{\eta
}\right)  \right)  ^{-1}\left(  \underset{k=0}{\overset{\infty}{\sum}}\left(
\frac{1}{4}\right)  ^{k}\left(  \frac{\eta-\frac{1}{\eta}}{\eta+\frac{1}{\eta
}}\right)  ^{k}H_{k}(x_{0})H_{l+k}(x_{1})\right)  .\label{L4}%
\end{equation}
Wave function (\ref{fi1}) describes then extended quantum object which
dynamical features and space-time extensions are linked through the light-cone
skeleton structure.

The two separate factors that constitute wave function (\ref{fi1}) describe
respectively kinetic and potential excitations. \ The ``oscillatory logic''
may suggest, however, that both kinds of excitations should be arranged in
``oscillatory order'', i.e. kinetic (potential) excitation should follow the
potential (kinematical) one and so on. Since $P_{\mu}P^{\mu}=m_{0}^{2}c^{2},$
the kinetic excitations provide us a picture of running plane waves carrying
out the four-momentum $P_{\mu}$ , or alternatively, the image of
point-particle endowed with the same four-momentum $P_{\mu}$. One needs to
keep in mind, however, that the area of plane waves propagation is limited to
the space-time region embraced basically by mentioned light-cone skeleton's
cover. On the other hand, the same light-cone skeleton provides us the
space-time extensions of the potential state. But this in turn mens that the
second factor of wave function (\ref{fi1}) describes the particle in the form
of extended material object moving through the space, as shown on Fig.6b. The
wave function (\ref{fi1}) provides then a new picture of particle-wave
duality, as well as, a new description of particle motion. According to this
the particle movement cannot be considered smooth but, as depicted on
Fig.5\textbf{, }must take a form of \textit{oscillatory}-like\textit{-motion}.

\subsection{The issue of uncertainty}

The outlined above peculiar kind of motion inevitable must be accompanied by
related energy and momentum fluctuations. To estimate the range of these
fluctuations one finds the light-cone skeleton structure to become very useful
again. It seems advisable, however, to recollect first the original concept of
Heisenberg uncertainties referred to quantum measurement process.

Alongside the postulates of Planck (\ref{du2}) and de Broglie (\ref{du1}%
)\ there are also Heisenberg uncertainty principles
\begin{equation}
\Delta x\Delta p\simeq h,\label{du3}%
\end{equation}
and
\begin{equation}
\Delta\mathcal{E}\Delta\tau\simeq h,\label{du4}%
\end{equation}
that constitute the basis of Schr\"{o}dinger quantum mechanics. It is widely
believed that formula (\ref{du3})\ describes relationship between the
uncertainties of particle momentum $\Delta p$ and position $\Delta x$ in a
measurement where both quantities are to be established at the same physical
process. Similarly, if the subject of measurement is the particle energy, then
related to this uncertainties of energy $\Delta\mathcal{E}$ \ and time
$\Delta\tau$ (needed for such measurement) are assumed to be in relation
(\ref{du4}). Seemingly both Heisenberg expressions have little to do with pure
relativistic approach and it seems rather unlikely to incorporate them into
rigid relativistic framework in a self-consistent way. Nevertheless, it turns
out to be feasible. There are however two essential obstacles that need to be
pointed out first.

The first remark concerns the underlying algebraic basis of \ both uncertainty
relations. Since the position and momentum are the $q-numbers$, i.e. position
and momentum have operator representatives $\widehat{x}$ and $\widehat{p}$ for
which it holds
\begin{equation}
\left[  \widehat{x},\widehat{p}\right]  =i\hbar,\label{uc14}%
\end{equation}
relation (\ref{uc14})\ is a $q-number$ uncertainty relation. In contrary to
this, as noticed by Dirac \cite{Dir27}, the (vital) time is (only) a
$c-number$, which mens that there is no Hilbert space associated to the time
variable \cite{Blanc82}. As a result it must occur
\begin{equation}
\left[  t,\widehat{H}\right]  =0,\label{com2}%
\end{equation}
where $\widehat{H}$ is energy operator or a Hamiltonian. Relation (\ref{com2})
then is a $c-number$ uncertainty relation. So, the emerging problem is that
Heisenberg relations brought directly on the relativistic ground forces the
Lorentz covariant description to deal with a mixture of $q-numbers$ and
$c-numbers,$ which, as also noticed by Dirac \cite{Dir27}, cannot be
consistent with special relativity.

The problem might become much simpler, however, if relations (\ref{du3}) and
(\ref{du4})\ are to be considered not in terms of $q-$ and $c-numbers$ but in
terms of wave packet average width-spreads. Let us then consider relation
(\ref{du3})\ and assume that it just relates the widths of wave packet spatial
distribution $\Delta x$ and longitudinal momentum distribution $\Delta p$.
However, if $\Delta x$ and $\Delta p$ are the quantities of Minkowski frame
(as it is commonly thought), then the boost transformation turns the relation
(\ref{du3}) into another one
\[
\Delta x\Delta p\simeq h\cdot sinh2\xi,
\]
which, of course, violates Heisenberg principle original form, thereby its
universal character originally assumed \cite{Kimh},\cite{Kim98}.

To overcome both mentioned difficulties Kim and Noz \cite{Kbook} showed up
that proper formulation of uncertainty principles provide canonically
conjugated Fourier components of wave function (\ref{b2}) i.e. the wave
function described in the photonic frame. Indeed, under such circumstances
relations (\ref{du3}) and (\ref{du4}) become the light-cone frame relations.
As a result they become Lorentz-boost form invariants, thereby become also the
integral part of the relativistic approach. The example of quantum object
spread out on light-cone skeleton structure illustrates this idea too.

The light-cone skeletons shown on Figs.4a\textbf{ }and 4b\textbf{,} are the
skeletons of the same quantum object but considered respectively in position
and momentum space. So that they are, of course, the mutually conjugated
structures. In order to emphasize that relations (\ref{du3}) and (\ref{du4})
should be regarded as the light-cone ones let us apply a symbolic
substitution
\begin{equation}
\Delta x\rightarrow\Delta\chi_{0},\text{ \ }\Delta p\rightarrow\Delta\pi
_{1},\label{v1}%
\end{equation}
and similarly
\begin{equation}
c\Delta\tau\rightarrow\Delta\chi_{0},\text{ \ }\frac{\Delta\mathcal{E}}%
{c}\rightarrow\Delta\pi_{0}.\label{v2}%
\end{equation}
As a result one finds that relations (\ref{du3}) and (\ref{du4}), considered
as the photonic frame uncertainty relations, become automatically fulfilled if
uncertainty values (\ref{v1}) and (\ref{v2}) become identified with the
light-cone skeleton extensions, i.e. when
\begin{equation}
\Delta\chi_{0}=\lambda_{0},\text{ }\Delta\chi_{1}=\lambda_{1},\text{ }%
\Delta\pi_{0}=\frac{h}{\lambda_{0}}\text{\ \ and }\Delta\pi_{1}=\frac
{h}{\lambda_{1}}.\label{SK1}%
\end{equation}

In order to return to the issue of oscillatory motion let us also answer the
question how the light-cone uncertainty relations
\begin{equation}
\Delta\chi_{0}\Delta\pi_{0}\simeq h\text{ \ \ and \ \ }\Delta\chi_{1}\Delta
\pi_{1}\simeq h,\label{LCun}%
\end{equation}
appear to the observer rest frame. Let us then recollect that effectively
observed particle dynamical and kinematical features, as it come from
(\ref{bi2}) and (\ref{lc1}) are those of the Minkowski frame. So, the same
must concern the energy and momentum fluctuations, which, then, must be
determined by
\begin{equation}
\left(
\begin{array}
[c]{c}%
\frac{\Delta p_{0}}{2}\\
\frac{\Delta p_{1}}{2}%
\end{array}
\right)  =\left(
\begin{array}
[c]{cc}%
\frac{1}{\sqrt{2}} & \frac{-1}{\sqrt{2}}\\
\frac{1}{\sqrt{2}} & \frac{1}{\sqrt{2}}%
\end{array}
\right)  \left(
\begin{array}
[c]{c}%
\frac{\Delta\pi_{0}}{2}\\
\frac{-\Delta\pi_{1}}{2}%
\end{array}
\right)  .\label{MKun}%
\end{equation}
Thus, for the given values of light-cone momenta (\ref{SK1}) one obtains that
energy and momentum fluctuations are
\begin{equation}
\Delta E=E\text{ \ \ and\ \ \ }\Delta p=p.\label{EnFl}%
\end{equation}
So, for the particle of model shape one obtains that its energy and
energy-fluctuations, as well as, momentum and momentum-fluctuations are the
same. This just explains the occurrence of (perfect) transitions between
kinetic and potential excitations, thereby oscillatory character of particle movement.

\subsection{Heisenberg-Witten vs. Kim-Noz uncertainty relations}

The broad issue of duality considered within the framework of string theory
presumably is one of the most exploring ideas in contemporary theoretical
physics. Even though the difficulty level even of quite simple string theory
analysis exceeds much that of given in this paper, it seems justified to
indicate some basic similarities that occur between the string theory and
covariant harmonic oscillator approach. The first is that in the both cases
one deals with space-time extended quantum objects instead of (field theory)
point-particle image. Secondly, it turns out that the source of particle mass
might be the \textit{string vibrations, }so that the concept of mass no longer
seem to be elementary. And finally, for the purpose of this paper, it is
enough to indicate one more common feature of the both approaches. Supported
by advanced calculations of Gross and Mende \cite{GM} and proposed in the
context of duality in string theory, generalized Heisenberg uncertainty
relation of Witten \cite{Witten}, is found to be exactly the one of Kim and
Noz (\ref{LCun}), however, being written at the Minkowski frame. It is
worthwhile to take a closer look at this quite elementary issue.

Let us consider again Eq. (\ref{lc1}) and focus on very time-like separation
$\Delta x_{0}.$ Due to (\ref{def3}), which allows us to express $\Delta x_{0}
$ by means of light-cone skeleton extensions $\lambda_{1}$ and $\lambda_{2},$
one obtains
\begin{equation}
\Delta x_{0}=\frac{1}{2}\lambda_{0}+\frac{1}{2}\lambda_{1}.\label{uc20}%
\end{equation}
Expressed in a similar way the energy fluctuations (\ref{MKun}) must take the
form
\begin{equation}
\Delta E=\frac{1}{2}\frac{hc}{\lambda_{0}}+\frac{1}{2}\frac{hc}{\lambda_{1}%
}.\label{uc21}%
\end{equation}
Eqs. (\ref{uc20}) and (\ref{uc21}) transparently reveal a dependence between
quantum object space-time extensions and its dynamical features. According to
(\ref{def3}) if particle stay at rest ($\eta=1$) both terms in both
expressions (\ref{uc20}) and (\ref{uc21}) contribute the same. In such case
the particle size is $\lambda_{C}$,\ which corresponds to $\Delta E=E=mc^{2}.
$ However, along with energy increase ($\eta\nearrow$, cf. formulas
(\ref{bi1}) and (\ref{bi2})) only one term of each of the formulas
(\ref{uc20}) and (\ref{uc21}) becomes dominant. Thus, even though the size of
the whole quantum object grows ($\Delta x_{0}$ $\sim\lambda_{0}$), its energy
fluctuations have still Heisenberg-like form, $\Delta E\Delta t\sim h$, where
$\Delta t=\lambda_{1}/c.$ Furthermore, since the energy increase induces
$\lambda_{1}$ decrease, dynamical consequences resulting from that make that
point-particle picture is to become more accurate too. Nevertheless, as it
comes from the discussion devoted to kinetic excitations, enlarged extensions
of quantum object indicate also the space-time area where the
``point-particle'' can be found. In other words, the formulas (\ref{uc20}) and
(\ref{uc21}), in simple quantum-mechanical manner, describe a new emerging
picture of particle wave-duality. But this is exactly the same what does
Witten formula predict. Indeed, by making use of property $\lambda_{0}%
\lambda_{1}=\lambda_{C}^{2}=(h/mc)^{2},$ one easily finds that expression
(\ref{uc20}) turns out to be (almost) the Witten formula
\begin{equation}
\Delta x_{0}=\frac{1}{2}\frac{h}{\Delta\pi_{0}}+\alpha_{m}^{\prime}\frac{1}%
{2}\frac{\Delta\pi_{0}}{h},\label{uc4}%
\end{equation}
where $\alpha_{m}^{\prime}=(h/mc)^{2},$ and $\Delta\pi_{0}=\frac{h}%
{\lambda_{0}}$. Thus, to expose the (possibly crucial) role of minimal length
scale (which is assumed to be the order of Planck length $l_{p}=\sqrt
{hG/\left(  2\pi c^{3}\right)  }$ $\sim1.6\cdot10^{-35}m,$ and thus
$\alpha_{m}^{\prime}\geqslant\alpha^{\prime}\equiv\left(  l_{P}\right)  ^{2}%
$),\textit{\ }as well as, to write down the Witten formula already in its
exact form, due to (\ref{uc4}), one may estimate minimal size of quantum
object according to
\begin{equation}
\Delta x_{0}\geqslant\frac{1}{2}\frac{h}{\Delta\pi_{0}}+\alpha^{\prime}%
\frac{1}{2}\frac{\Delta\pi_{0}}{h}.\label{uc22}%
\end{equation}
So, as one would expect, the Planck length $l_{P}$ is the Lorentz invariant
indeed and, as it comes from above, there is no need to ``double'' relativity
issue \cite{AM} to support that statement. Nevertheless, a particle cannot be
seen as a material point but rather as an extended quantum object endowed with
internal space-time structure.

\section{Concluding remarks}

It is widely believed that the origin of special relativity is purely the
classical one. On the other hand, there is no doubt that there would be no
Einstein theory if the Maxwell equations and observations of Michelson and
Morley concerning the ether existence had become known first. The key point
is, however, that ``early observations'' of electromagnetic interactions are
thought to be the classical ones too. Even though that ``classical
electrodynamics'' blows up the framework of Newtonian physics, the name
``classical'' does not seem to be used improperly. Given that electromagnetic
field indeed is a true classical field (whatever it means) its quantum nature
was already recognized by Planck at the problem of blackbody radiation, in
1901, i.e. before the special relativity came up. What's more, the quantum
nature of electromagnetic field was confirmed by Einstein himself \ who
discovered the photoelectric effect in 1905. Nevertheless the well-know
Einstein's stubbornness against comprehension of physical word in terms of
quantum mechanics has been never overcame.

In 1935 Einstein, Podolsky and Rosen \cite{EPR} (EPR)\ in an effort to rescue
``locality'' and ``classical reality'' introduced their famous
Gedankenexperiment and proposed that quantum mechanics was incomplete. Let us
then remind the Einstein locality principle \cite{Ein2}: ``If $S_{1}$ and
$S_{2}$ are the two systems that have interacted in the past but now are
arbitrary distant, the real factual situation of system $S_{1}$ does not
depend on what is done with system $S_{2},$ which is spatially separated from
the former.''\ Then, the new class of models, called the Local
Hidden-Variables (LHV) models, appeared to describe statistical features of
quantum measurements as a result of underlying deterministic substructure. In
1964 Bell showed \cite{Bell64}, however, that all LHV models that provide the
results being in complete agreement with predictions of quantum mechanics do
not obey the principle of locality, or, in other words, if substructures of
LHV models appear to be \textit{truly local}, their predictions must differ
from those of quantum mechanics.

At early seventies, with the aid of Bell (inequalities) and available new
experimental techniques, a new era of Gedankenexperiments begun and despite of
EPR expectations the results it has provided have testified strongly against
the classical ideology. The violation of Bell-inequalities were observed in a
wide range of Gedankenexperiments based mainly on two-photons correlations
measurements such as: polarization correlations \cite{ss}, energy and time
correlations (followed by experimental proposal of Franson \cite{Fr})
\cite{Et} , or phase and momentum correlations \cite{pm}.

Additionally, it is worthwhile to single out three more experimental
observations that disclose unlocal properties of quantum states in very
spectacular way, namely: the Franson and Potocki observation of ``Single
photon interference over large (45m) distances'' \cite{FP}, G. Weiss et al.
\cite{Weiss} ``Violation of Bell's inequality under strict Einstein locality
conditions'', and W. Tittel at al. \cite{Tittel} ``Experimental demonstration
of quantum correlations over more then 10 km.''

The era of Gedankenexperiments is basically finished and its heritage was
taken over by Quantum Teleportation already originated over ten years ago
\cite{tele}. Although above mentioned experimental results clearly indicate
that tight keeping on Einstein's locality idea contradicts the sober view, the
issue of Lorentz symmetry, so far, has been never regarded in terms of the
quantum approach seriously. Furthermore, even from the very theoretical point
of view, one finds that Maxwell equations do occupy exactly the same position
in quantum field theory approach as the equations of Dirac and Klein-Gordon
do. Of course, these two material fields, until quantized, play the role of
classical fields too. Nevertheless, devoid of the context of quantum mechanics
they mean nothing. In the case of electromagnetic field and Lorentz symmetry
the situation, presumably, is very similar.

And finally let us ask the fundamental question: is that because of Einstein's
time relativity idea looking so attractive, any alternative approach to
``relativity issue'' is simply out of the question? The honest answer is,
perhaps, as hard as solutions of a few challenging physical problems. The
author of the parer is aware of its simplicity, thereby far from the belief
that presented now non-orthodox approach might be called satisfied. It seems,
however, that even more important then any field theory analysis is to realize
first how substantial and ``positive flooding'' consequences of departure form
orthodox relativity perception might be. The main results of analysis already
started in \cite{PK267} and being continued now are summarized below in the
form of concise comparison between some consequences resulting from orthodox
and non-orthodox views.%

\begin{align*}
&
\begin{array}
[c]{cc}%
\text{ORTHODOX \ \ VIEW \ \ \ \ \ \ } & \text{NON-ORTHODOX VIEW\ }%
\end{array}
\text{\ \ \ \ \ \ \ \ \ \ \ \ \ \ \ \ \ \ \ \ \ \ \ \ \ \ \ \ }\\
&  \text{ \ \ \ \ \ \ \ \ \ \ \textbf{General Meaning of Special Relativity }%
}\\
&
\begin{array}
[c]{cc}%
\text{The Classical theory \ \ \ \ \ \ \ \ } & \text{The Tool of Quantum
Mechanics\ \ \ \ \ }%
\end{array}
\\
&  \text{ \ \ \ \ \ \ \ \ \ \ \textbf{The Special Relativity Source}}\\
&
\begin{array}
[c]{cc}%
\text{Relativity of inetrial frames} & \text{Freedom of choice of comparison
scale}%
\end{array}
\text{ }\\
&  \text{
\ \ \ \ \ \ \ \ \ \ \ \ \ \ \ \ \ \ \ \ \ \ \ \ \ \ \ \ \ \ \ \ \ \ \ \ \ \ \ \ maintained
within the framework of preferred}\\
&  \text{ }%
\ \ \ \ \ \ \ \ \ \ \ \ \ \ \ \ \ \ \ \ \ \ \ \ \ \ \ \ \ \ \ \ \ \ \ \ \ \ \ \ \text{frame
quantum description}\\
&  \text{ \ \ \ \ \ \ \ \ \ \textbf{The Emerging Particle Image}}\\
&
\begin{array}
[c]{cc}%
\text{Point-like \ \ \ \ \ \ \ \ \ \ \ \ \ \ \ \ \ \ \ \ \ \ \ } &
\text{Extended quantum object endowed with }%
\end{array}
\\
&  \text{
\ \ \ \ \ \ \ \ \ \ \ \ \ \ \ \ \ \ \ \ \ \ \ \ \ \ \ \ \ \ \ \ \ \ \ \ \ \ \ \ internal
space-time structure}\\
&  \text{\ }\ \ \ \ \ \ \ \ \ \ \text{\textbf{Basic Relativity Predictions}}\\
&
\begin{array}
[c]{cc}%
\text{Relativity of time and \ \ \ \ \ \ } & \text{Space-time deformations of
quauntum }%
\end{array}
\text{\ \ \ }\\
&
\begin{array}
[c]{cc}%
\text{lenght maesure \ \ \ \ \ \ \ \ \ \ \ \ \ \ \ } & \text{object internal
structure. }%
\end{array}
\\
&  \text{
\ \ \ \ \ \ \ \ \ \ \ \ \ \ \ \ \ \ \ \ \ \ \ \ \ \ \ \ \ \ \ \ \ \ \ \ \ \ \ \ Time
and length measures remain intact }%
\end{align*}

I would like to thank prof. W. W\'{o}jcik, prof. B. Kozarzewski and prof. A.
B\"{o}hm for the financial support that allowed me to participate on the
Second Feynman Festival conference held at the University of Maryland (August
2004, Collage Park, MD. U.S.A.)

\end{document}